\documentclass[runningheads]{llncs}
\usepackage[T1]{fontenc}
\usepackage{graphicx}
\usepackage{booktabs}
\usepackage{amsmath}
\usepackage{amsfonts}
\usepackage{xspace}
\usepackage{xcolor}
\usepackage{tikz}
\usepackage{subfig}
\usepackage[hyphens]{url}
\usetikzlibrary{arrows}
\usetikzlibrary{positioning}

\newcommand\ie{i.\,e.\xspace}
\newcommand\eg{e.\,g.\xspace}

\begin{document}
\title{An Empirical Evaluation of Predicted Outcomes as Explanations in Human-AI Decision-Making}
\titlerunning{Predicted Outcomes in Human-AI Decision-Making}

\author{Johannes Jakubik \and
Jakob Schöffer \and
Vincent Hoge \and 
Michael Vössing \and 
Niklas Kühl 
}
\authorrunning{J. Jakubik et al.}

\institute{
Karlsruhe Institute of Technology (KIT), Germany\
\email{\{johannes.jakubik,jakob.schoeffer,michael.voessing,niklas.kuehl\}@kit.edu}
\email{vincent.hoge@alumni.kit.edu}}
\maketitle

\begin{abstract}
In this work, we empirically examine human-AI decision-making in the presence of explanations based on predicted outcomes. 
This type of explanation provides a human decision-maker with expected consequences for each decision alternative at inference time---where the predicted outcomes are typically measured in a problem-specific unit (e.g., profit in U.S. dollars).
We conducted a pilot study in the context of peer-to-peer lending to assess the effects of providing predicted outcomes as explanations to lay study participants.
Our preliminary findings suggest that people's reliance on AI recommendations increases compared to cases where no explanation or feature-based explanations are provided, especially when the AI recommendations are \emph{incorrect}.
This results in a hampered ability to distinguish correct from incorrect AI recommendations, which can ultimately affect decision quality in a negative way.
\end{abstract}

\section{Introduction}\label{sec:introduction}
In real-world decision-making, human decision-makers are confronted with a range of available decision options with diverging future outcomes. 
For this reason, several approaches in the field of prescriptive AI emerged to support human decision-makers by not only recommending a decision option but also quantifying the predicted outcomes of \emph{all} available decision options (e.g., expected profit in U.S. dollars). 
For decades, these approaches have been leveraged in a range of real-world high-stakes decision-making scenarios, such as in medical and healthcare \cite{Bertsimas2019OptimalTrees,bertsimas2020prescriptive,wang2019prescriptive}, financial \cite{khatri2019analytics}, manufacturing \cite{ansari2019prima,matyas2017procedural}, or strategic management~\cite{postma2005improve} domains. 
In line with this, large tech companies such as GE\footnote{\url{https://www.cio.com/article/244505/ge-pitney-bowes-team-up-on-predictive-and-prescriptive-analytics.html} (last accessed July 27, 2022)}, IBM\footnote{\url{https://www.ibm.com/analytics/prescriptive-analytics} (last accessed July 27, 2022)}, or Microsoft\footnote{\url{https://appsource.microsoft.com/en-us/product/web-apps/river-logic.riverlogic_analytics?tab=overview} (last accessed July 27, 2022)} have been investing in prescriptive AI. 
However, there is a lack of empirical analyses on the effects of these predicted outcomes on human-AI decision-making in general. 
We hypothesize that presenting predicted outcomes of decision options to human decision-makers can influence their reliance on AI recommendations (\eg, a human decision-maker might refrain from choosing decision options with a negative predicted outcome and, therefore, follow the AI even when the AI is incorrect). 
Hence, this work sets out to empirically assess the influence of predicted outcomes on the performance of human-AI decision-making in general and on humans' reliance on AI recommendations specifically.

Predicted outcomes inform human decision-makers why a certain decision option is recommended instead of an alternative one (e.g., ``do \emph{not} lend money to this person because the predicted financial return of lending the money is negative''). This is in line with the definition of \textit{why not} explanations~\cite{lim2019these}. \textit{Why not} explanations provide information on why an inferred recommendation and not an alternative one was produced. Hence, these explanations are \textit{contrastive} in the sense that they allow for a pairwise comparison between the inferred and an alternative recommendation (see, e.g., \cite{Miller2019}). Typically, \textit{why not} explanations take into account current input values to inform human decision-makers why a specific decision option is recommended instead of alternative options. In contrast to this, predicted outcomes explain why a decision option is recommended based on expected future returns of all decision options, which are inferred by the model together with a decision recommendation. Thus, instead of descriptive information about the model input, predicted outcomes explain decision recommendations based on expected future consequences. This characteristic makes studying predicted outcomes of decision alternatives especially relevant for the XAI community.

The results of our in-progress work indicate that study participants tend to follow AI recommendations more often when these recommendations are supplemented with predicted outcomes, as compared to other conditions where they are given no explanation or feature-based explanations.
This effect is particularly pronounced when AI recommendations are incorrect---a phenomenon commonly referred to as \emph{over-reliance}.
Importantly, when the AI recommendation is supplemented with predicted outcomes, we observe a tendency towards a reduced ability of study participants to distinguish between correct and incorrect AI recommendations. 
Thus, our preliminary findings suggest that using predicted outcomes as explanations can be detrimental to human-AI decision-making.

\section{Related work}\label{sec:Background} 
In the following subsections, we present related literature on XAI and reliance in human-AI decision-making.

\subsection{Explainable AI}\label{sec:XAI}
AI algorithms can provide powerful decision support and have 
already become ubiquitous in many domains~\cite{kuncel2014hiring,townson2020ai}. 
Problematically, many AI algorithms are opaque, which means it is difficult for users to gain insight into the internal processes and to understand why the AI suggests a specific decision \cite{Adadi2018}. XAI is concerned with making AI-based systems more transparent by providing explanations for black-box models \cite{guidotti2018survey} or by using interpretable machine learning models \cite{rudin2019stop}.
Transparency is widely assumed to improve human-AI decision-making by enabling users to detect and correct errors of the AI and by ensuring that AI decisions are fair \cite{Binns2018ItsDecisions,das2020opportunities,Dodge2019ExplainingJudgment,vossingDesigningTransparencyEffective2022}. 
Additionally, there is a demand for explanations to comply with legislation, for example, the EU General Data Protection Regulation (GDPR).

Despite these claims, recent research shows that XAI does not necessarily improve human-AI decision-making over cases where no explanations are provided \cite{Alufaisan2020DoesDecision-Making,Green2019TheMaking,metaanalysis}. 
Even worse, \cite{Poursabzi-Sangdeh2021ManipulatingInterpretability} find that providing people with an interpretable model can result in less accurate predictions.
Yet, some studies show better human-AI decision performance when AI predictions are supplemented with explanations, compared to the performance when only predictions are provided (e.g., \cite{Bucinca2020ProxySystems,Lai2019}).

Common XAI methods are feature-based and rule-based explanation approaches \cite{Alufaisan2020DoesDecision-Making}. 
Feature-based models provide the most important features responsible for the output of the machine learning algorithm and its associated weights.
Rule-based explanations output \emph{if-then-else} rules which state the decision boundary between the given and contrasting predictions \cite{Alufaisan2020DoesDecision-Making,vanderWaa2021}.
Since feature-based explanations are among the most commonly employed XAI approaches, we include them in our study as a baseline.

\subsection{Reliance in human-AI decision-making}\label{sec:Reliance}
Reliance is defined as a behavior \cite{lee2004trust} that, in the context of human-AI decision-making, is referred to as following an AI recommendation \cite{schemmer2022should,vereschak2021evaluate}.
However, it is not always beneficial to rely on AI recommendations, given that AI may be imperfect and may provide incorrect recommendations.
People following incorrect AI recommendations---also referred to as \emph{over-reliance}---is a major issue that can inhibit human-AI complementarity  \cite{Bucinca2021}.
To establish human-AI complementarity, humans need to \textit{appropriately} rely on AI recommendations, meaning people must be able to distinguish correct and incorrect AI recommendations and act upon that differentiation  \cite{schemmer2022should,schoeffer2022relationship}.

Prior findings regarding the effects of XAI on reliance are inconclusive but show a tendency towards increased over-reliance. 
For example, \cite{vanderWaa2021} discovered an increased reliance for example- and rule-based explanations---also on incorrect AI recommendations.
In the study of \cite{Lai2019}, study participants followed AI recommendations significantly more often when provided with example- and feature-based explanations, even if they contained random content.
\cite{Poursabzi-Sangdeh2021ManipulatingInterpretability} observed that study participants supplemented with an interpretable model were less able to detect mistakes of the model compared to study participants provided with a black-box model---likely due to information overload. 
Besides information overload, over-reliance in human-AI decision-making may be caused by, for example, heuristic decision-making \cite{Bucinca2021}. 
The authors of the study hypothesize that people develop heuristics about the overall competence of the AI~\cite{Bucinca2021}.
In this context, explanations are interpreted as a general sign of competence of the AI, which then leads people to follow AI recommendations without thoroughly vetting them.

In prior XAI research, many approaches for explaining AI systems have been developed and evaluated with respect to their effects on human-AI complementarity. 
However, the effects of predicted outcomes as explanations have not been studied yet. 
As predicted outcomes play an important role in scenario analyses and high-stakes decision-making (\eg, medical \cite{Bertsimas2019OptimalTrees}, financial \cite{khatri2019analytics}, or strategic management~\cite{postma2005improve} domains), we aim to better understand the effects of such explanations on human-AI decision-making.

\section{On the relationship of reliance and human-AI decision-making accuracy}

In the following, we discuss the general influence of reliance $\mathbf{r}$ on human-AI decision-making accuracy $\mathcal{A}$ for a given AI performance. For this, we define reliance as the proportion to which people follow AI recommendations in human-AI decision-making. Over-reliance then refers to a situation in which people follow the AI not only in cases when the AI recommendation is correct but even when the given recommendation is incorrect. We define the opposite phenomenon as \emph{under-reliance}. We then model the human-AI decision-making accuracy as a function of reliance $\mathcal{A}(\mathbf{r})$. We observe that for $\mathbf{r}\longrightarrow 1$, the human-AI decision-making performance will converge to the accuracy of the AI. For $\mathbf{r} \in (0, 1)$, the human-AI decision-making accuracy ranges in an interval $\mathcal{A}(\mathbf{r})=[min, max]$ that indicates the minimum and maximum of the possible human-AI accuracy. Imagine, for example, an AI accuracy of 66.7\% and a reliance of $\mathbf{r}=66.7\%$. People may correct the AI in all cases where the AI recommendation is incorrect, which would result in a human-AI decision-making accuracy of 100\%. However, when people incorrectly override the AI in all cases where the AI recommendation is correct, the resulting human-AI decision-making accuracy would be 33.3\%, \ie, $\mathcal{A}(\mathbf{r}=66.7\%)=[33.3\%, 100\%]$. We observe that the interval of possible human-AI decision-making accuracy is largest when $\mathbf{r}$ is equal to the AI accuracy and becomes smaller with increasing $\mathbf{r}$ (\eg, $\mathcal{A}(\mathbf{r}=90\%)=[43.3\%, 77.7\%]$), finally converging to the accuracy of the AI. 

\section{Study design}\label{sec:StudyDesign}
In this section, we first formulate our research hypotheses.
Then, we outline the use case and dataset chosen for this study, and we address technical preliminaries.
Finally, we introduce our experimental design and the process of recruiting study participants.

\subsection{Hypotheses}\label{sec:Hypotheses}
Prior research already discovered that XAI can have effects on reliance. 
While many studies report XAI leading to over-reliance \cite{schemmer2022influence}, the effect demands further investigation \cite{metaanalysis}. 
The results of multiple studies remain inconclusive, some pointing towards over-reliance \cite{Bucinca2021,schoeffer2022relationship}, some to under-reliance \cite{nourani2021anchoring,schemmer2022should}. 
When it comes to the effects of \textit{predicted outcomes}, multiple researchers raise the question on their influence on reliance and accuracy \cite{antoniadi2021current,mueller2019explanation}---with some suspecting a trend towards over-reliance \cite{keane2021if,naiseh4098528different}. 
Thus, we conducted an exploratory pilot study to examine the effects of predicted outcomes as explanations on human-AI decision accuracy and human reliance on AI recommendations. 

\begin{itemize}
\item[\textbf{H1}] People provided with predicted outcomes as explanation follow an AI recommendation more often than people provided with an AI recommendation without explanation.
\end{itemize}

We further hypothesize that on average and for a certain level of reliance, the empirical human-AI decision-making performance will be close to the mean value $\overline{\mathcal{A}(\mathbf{r})}$ of the interval $\mathcal{A}(\mathbf{r})=[min, max]$, as introduced previously. Thus, even when people follow the AI in too many cases (\ie, over-reliance), we hypothesize that the human-AI decision-making accuracy is still given by $\overline{\mathcal{A}(\mathbf{r})}$.

\begin{itemize}
\item[\textbf{H2}] 
The empirical human-AI decision-making accuracy is close to the mean value $\overline{\mathcal{A}(\mathbf{r})}$ of the theoretical function $\mathcal{A}(\mathbf{r})$. 
\end{itemize}

For many use cases, human-AI decision-making represents a special form of decision-making under risk, as defined by \cite{kahneman1979prospect}. When predicted outcomes as explanations come into play (e.g., in terms of potential future consequences of the available decision options), we follow prospect theory in assuming that \textit{losses loom larger than gains}. We thus expect that people tend to follow AI recommendations supplemented by negative predicted outcomes more often in order to avoid potential losses in the future. 

\begin{itemize}
\item[\textbf{H3}] People follow AI recommendations supplemented by predicted outcomes more often when the predicted outcomes are negative.
\end{itemize}

\subsection{Preliminaries}\label{sec:UseCase}

\paragraph{Use case}
For our study, we train the prescriptive AI on a real-world dataset.
We use a publicly available dataset on peer-to-peer loans from the financial company Lending Club\footnote{\url{https://www.kaggle.com/datasets/wordsforthewise/lending-club} (last accessed July 27, 2022)}.
Lending scenarios have been frequently studied in prior XAI user studies (\eg, \cite{Confalonieri2021,Green2019TheMaking}) and constitute a relevant use case for prescriptive AI. 
The Lending Club dataset comprises real-world observations from a peer-to-peer lending platform that enabled individuals to lend money to others.
As borrowers potentially fail to completely pay back the owed money, it is essential for lenders to accurately assess the risk of defaulting. 
In this scenario, prescriptive AI could provide valuable decision support.

\paragraph{Dataset}
Our dataset contains 2,260,701 loans issued from 2007 until the end of 2018.
We only consider loans that were either fully paid off or defaulted, resulting in a dataset of 1,331,863 loans.
About 80\% of these loans were fully repaid.
The dataset contains 150 features and the label whether the borrower defaulted on the loan or not.
To achieve a reasonable task complexity for human-AI decision-making, we limit the data to 6 features: \textit{
borrower’s monthly income, FICO credit score, interest rate, loan amount, number of months to pay off the loan,} and the \textit{amount of each monthly installment}.
This selection of features from the Lending Club dataset is consistent with related literature (\eg, \cite{Green2019TheMaking}).

\paragraph{Technical preliminaries}
Prescriptive AI methods recommend (\ie, prescribe) the best option among a set of available decision alternatives---typically by maximizing the predicted outcome of the set of available decision options. 
In our case, we utilize prescriptive trees as an exemplary prescriptive AI to calculate predicted outcomes and the resulting AI recommendation \cite{Bertsimas2019OptimalTrees}. 
Several other approaches of prescriptive AI utilize predicted outcomes as well (\eg, \cite{bastani2020online,chen2022statistical}). 
Note, that prescriptive trees provide a range of additional measures designed for human experts to increase the interpretability of the prescriptive AI, which are not part of our study. 
A major challenge for decision-making in general (and, therefore, also for prescriptive AI), is that the true outcome can only be observed for the selected decision option in real-world use cases. 
Hence, outcomes of alternative decision options and the overall correct decision are unknown \cite{lakkaraju2017selective}. 
These unknown outcomes are often called \emph{counterfactuals}. 
The prescriptive AI is, therefore, trained for an accurate estimation of the counterfactual outcomes. 

In the following, we outline the technical approach behind several prescriptive AI. 
The prescriptive AI is trained on observational data $\{(x_i,y_i,z_i)\}_{i=1}^{n}$, including feature values $x_i \in \mathbb{R}^{d}$ of each observation $i$ with $d$-dimensional feature vectors, the assigned decision $z_i \in\{1, . . ., m\}$ and the corresponding outcome $y_i \in \mathbb{R}$ under the decision for $n \in \mathbb{N}$ realizations. 
For the accurate estimation of the counterfactual outcomes, the model aims at minimizing the squared prediction error for the observed data: $\sum_{i=1}^{n}\left(y_{i}-\hat{y}_{i}\left(z_{i}\right)\right)^{2}$.
Here, $\hat{y}_{i}(t)$ refers to the unknown outcome that would have been observed if decision $t$ had been chosen for sample $i$. 
The overall goal of the prescriptive AI is to simultaneously estimate counterfactual outcomes for \textit{all} decision options and to prescribe the option that optimizes the predicted outcome. 
Thus, in contrast to predictive AI, the prescriptive AI implicitly infers both predicted outcomes and a recommended decision option within a single model.

We evaluate the performance of the prescriptive AI by comparing the prescribed decision with the optimal 
decision based on synthetic ground truth, following, for example, \cite{Bertsimas2019OptimalTrees}.
The model achieves an accuracy of 85\% accompanied by an area under receiver operating characteristic (AUROC) score of 86\%. The model prescribes to lend money to the borrower for approximately 62\% of the instances.

\subsection{Experimental design}\label{sec:ExperimentalDesign}

The purpose of our study is to examine how supporting humans with predicted outcomes affects human-AI decision accuracy and the reliance of humans on AI recommendations.
Therefore, we conduct a scenario-based online experiment. In our experiment, we present loan applications to the study participants and ask them to decide whether to lend money to the applicant or not.
The study participants are assisted by AI recommendations and different types of explanations.
We use a between-subjects design with three experimental conditions as outlined in Table \ref{tab:conditions}. 
We utilize feature-based explanations as a baseline to better understand the effect sizes of explanations based on predicted outcomes. As the utilized prescriptive AI is tree-based, we follow \cite{Bertsimas2019OptimalTrees} and calculate the global feature importance. The importance of each feature is denoted by the total decrease in the loss function as a result of each split in the trees that include this feature. The resulting scores are normalized so that the feature importance sums to 100\%.

\begin{table*}[htbp]
    \centering
    \caption{Experimental conditions of our study design.}
    \resizebox{\textwidth}{!}{
    \begin{tabular}{p{6cm} p{7cm}}
        \toprule
        \bf Condition & \bf Explanation \\
        \midrule
        \bf AI without explanation & Study participants are provided only with an AI recommendation, not with predicted outcomes associated with the decision options.  \\
        \midrule
        \bf AI with predicted outcomes & Study participants are provided with an AI recommendation and, additionally, with the predicted outcomes for both decision options. \\
        \midrule
        \bf AI with feature-based explanation & The AI recommendation is shown to the study participants and, additionally, the feature importance scores calculated by the model. This condition represents a common XAI approach and therefore serves as a baseline. \\
        \bottomrule
    \end{tabular}
    }
\label{tab:conditions}
\end{table*}

\begin{figure*}[htb!]
 \centering
 \frame{\includegraphics[width=0.8\linewidth]{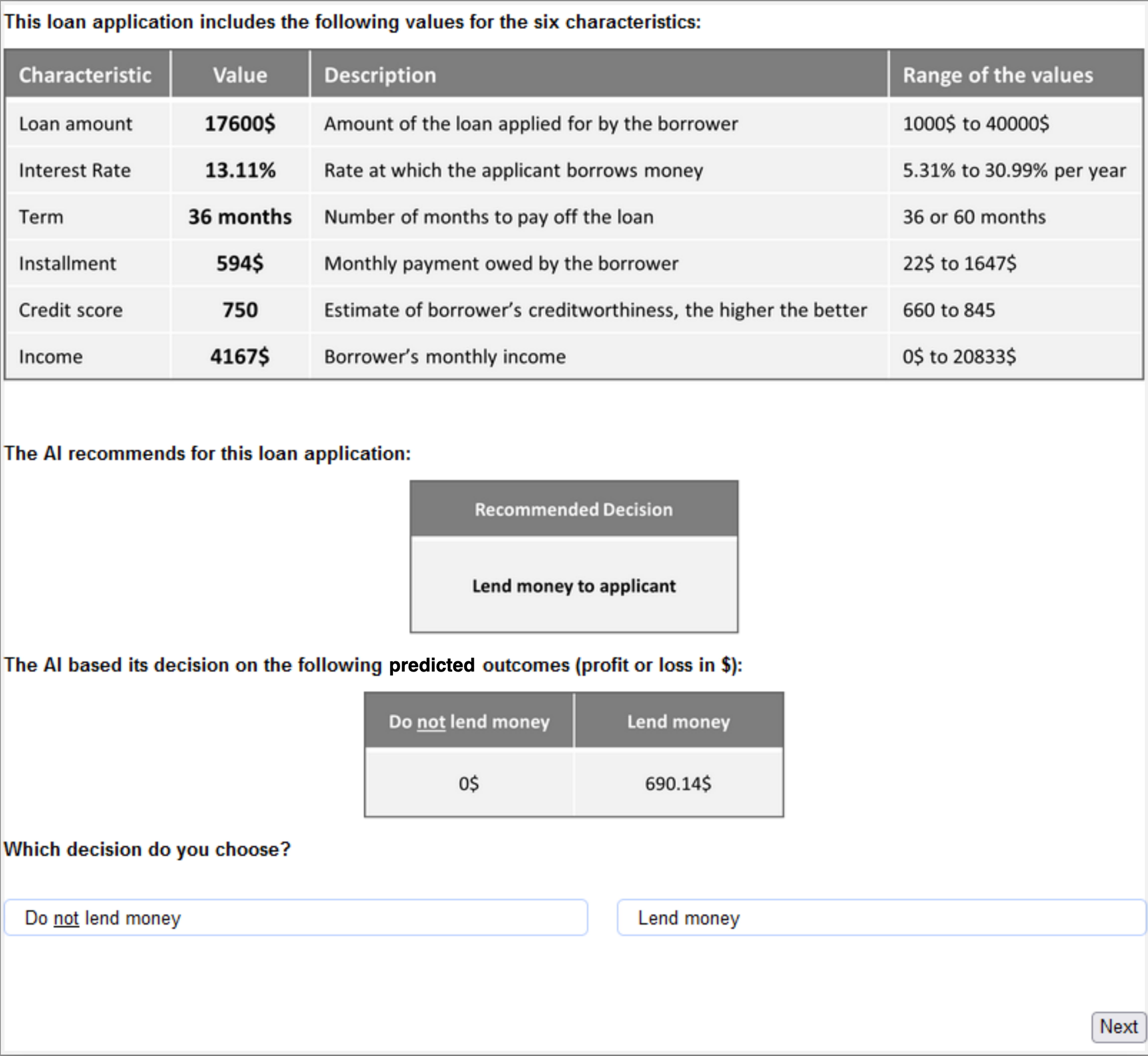}}
 \caption{Exemplary trial from our study presenting the task and relevant information in the \emph{AI with predicted outcomes} condition.}
 \label{fig:survey}
\end{figure*}

The study participants are randomly assigned to one of the conditions. In each condition, study participants are working on the same set of loan applications.
Each loan application is characterized by the 6 observational features.
A description of the features and the range of values (in the entire dataset) are displayed throughout the decision-making task (see Figure~\ref{fig:survey}).
By varying only the type of explanation, we can measure the effect of each treatment on the decision-making behavior.
Human-AI decision-making accuracy is measured by the percentage of instances where study participants select the correct decision option (\ie, the option the reward estimation suggests).
We quantify reliance by measuring the share of instances for which humans follow the AI recommendation. Over-reliance is given by the share of instances for which human decision-makers follow an \textit{incorrect} recommendation.

\begin{figure*}[htb!]
 \centering
 \frame{\includegraphics[width=0.8\linewidth]{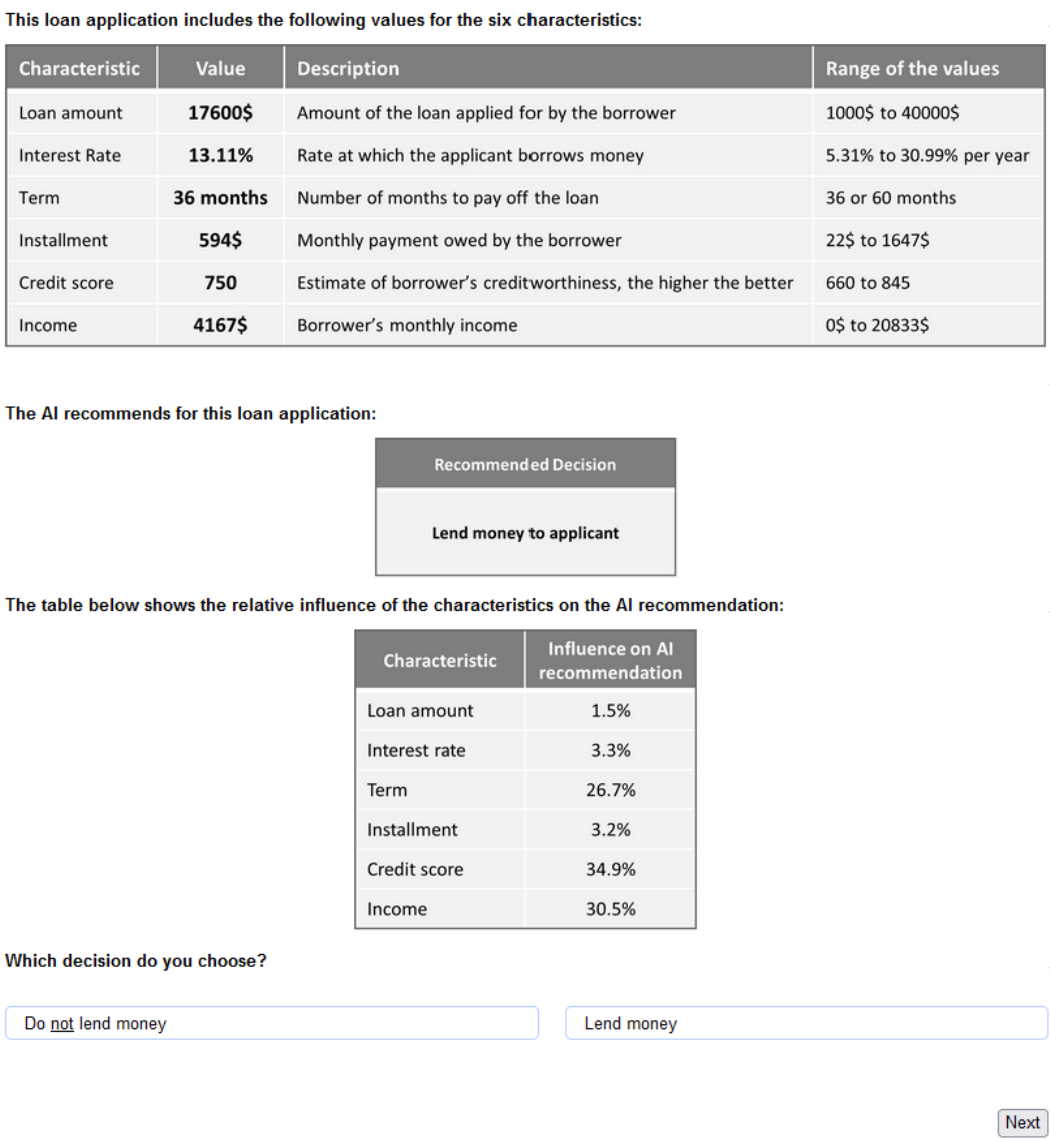}}
 \caption{Exemplary trial from our study presenting the task and relevant information in the \emph{AI with feature-based explanation} condition.}
 \label{fig:survey_feat_imp}
\end{figure*}

Our study includes a consent form followed by an introduction to the task, a training and testing phase, as well as questions about demographic information and proficiency in the fields of AI and lending. 
In the training phase, study participants are familiarized with the procedure of the experiment, the domain, and the AI recommendations.
The training phase consists of three randomly ordered trials. In each trial, study participants are shown the instructions for the task specific to the assigned condition, a loan application, the AI recommendation, and the corresponding explanation depending on the assigned condition (see Figure \ref{fig:survey} for an exemplary trial with predicted outcomes as explanations, and Figure \ref{fig:survey_feat_imp} with feature-based explanations).
The study participants must then choose whether they would lend money to the applicant. In the training phase, after submitting a decision, the study participants are informed about what would have been the correct decision. 
For the training phase, we randomly sample two loan applications where the model recommends the correct decision option and one application where the model is incorrect. 
Thus, the study participants learn that the AI recommendation could be incorrect.
We do not report results from this training phase.

In the testing phase, study participants decide on 12 loan applications. Similar to the training phase, the AI recommendation is correct for 8 loan applications and incorrect for the remaining 4 trials.
Thus, in our sampling, the AI recommendation is correct in 66.7\% of the cases. The cases where the AI recommendation is incorrect are composed of two trials where the AI \emph{incorrectly} recommends to give a loan, and two trials where the AI \emph{incorrectly} recommends to reject a loan application. The incorrect AI recommendations later allow us to determine whether study participants over-rely on the AI by following wrong AI recommendations.
The trials are then presented to the study participants in random order.
The procedure in the testing phase resembles the one in the training phase, except that we do not provide information on which decision would have been correct after study participants submit their decision. 
We collect the decisions of the study participants throughout the testing phase and later report our results based on the study participants' decisions. 

\subsection{Study participants}\label{sec:Participants}
We recruited 121 study participants via Prolific---a crowdworking platform for online research\footnote{\url{https://www.prolific.co/} (last accessed July 27, 2022)} \cite{palan2018prolific}. 
Study participants were not required to have explicit expertise in lending or loan applications to participate in our study. 
The study participants were randomly assigned to one of the three conditions.
Each study participant received a base payment of \$1.50 for completing the study.
As an incentive for study participants to do their best during the test phase, they were rewarded with an additional bonus payment of \$0.04 for each correct decision, resulting in a maximum total bonus of \$0.48.
The median time to complete the study was approximately 10 minutes. 

\section{Results}\label{sec:Results}
In this section, we report the results from our pilot study and analyze the effects of the different conditions on (a) the reliance of study participants on AI recommendations, and (b) human-AI decision-making performance.

\subsection{Reliance on AI recommendations}\label{sec:RelianceResults}
\begin{figure}[ht]
 \centering
 \includegraphics[width=0.6\linewidth]{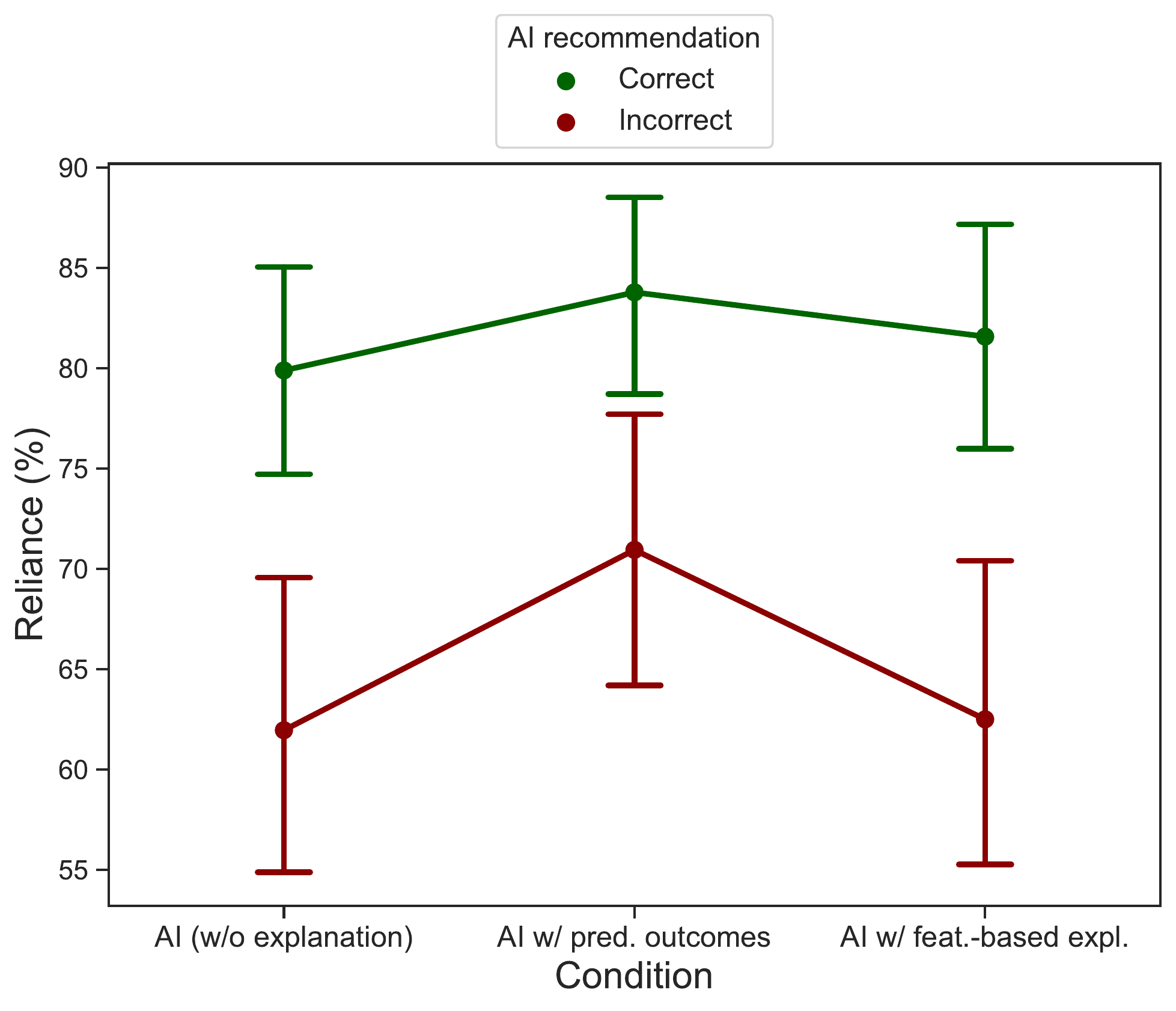}
 \caption{Reliance of study participants on correct and incorrect AI recommendations per condition. Error bars represent 95\% confidence intervals.}
 \label{fig:plot_point_reliance}
\end{figure}

As we cannot confirm the assumption of normality, we employ non-parametric Kruskal-Wallis tests \cite{kruskal1952use} to test for differences across the conditions in our experiment.
Subsequently, we conduct post-hoc pairwise comparisons between conditions by utilizing Bonferroni-corrected Mann-Whitney U tests \cite{mann1947test}.
Figure \ref{fig:plot_point_reliance} shows the reliance of study participants on correct and incorrect AI recommendations for each condition. 
First of all, study participants generally followed correct AI recommendations more often than incorrect AI recommendations ($p < 0.001$).
This also applies to each specific condition, where we find a significant difference in reliance on correct versus incorrect AI recommendations.
We infer from this that study participants were able to distinguish between correct and incorrect AI recommendations---even without explanations.

Importantly, our results in Figure~\ref{fig:plot_point_reliance} imply a difference between the over-reliance\footnote{Recall that we define \emph{over-reliance} as following \emph{incorrect} AI recommendations.} on AI recommendations without explanations ($mean= 62.0\%,\ std =  26.2\%$) and the over-reliance on AI recommendations with predicted outcomes as explanations ($mean = 70.9\%,\ std = 21.7\%$). 
This observation aligns with hypothesis \textbf{H1}.
However, due to the relatively small sample size in our pilot study, we cannot report statistical significance ($p = 0.14$). 
We further do not observe this tendency when comparing the over-reliance on AI recommendations without explanations with the over-reliance on AI recommendations supplemented with feature-based explanations ($mean = 62.5\%,\ std = 23.8\%$).

\begin{figure}[t]
 \centering
 \includegraphics[width=0.6\linewidth]{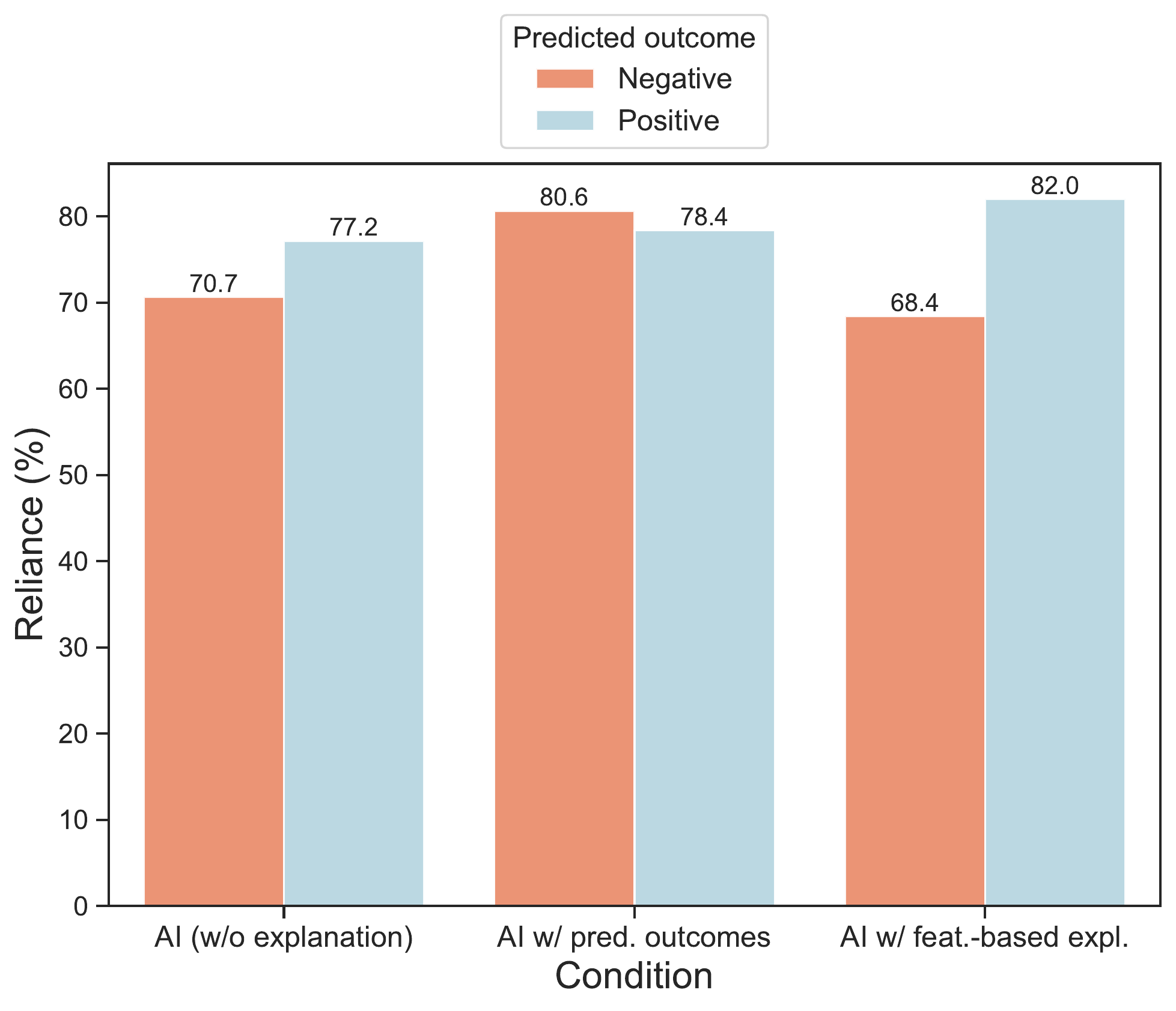}
 \caption{Reliance of study participants on AI recommendations with negative and positive predicted outcomes per condition.}
 \label{fig:plot_bar_reliance_outcomes}
\end{figure}

We additionally analyze the influence of positive and negative predicted outcomes on the reliance on AI recommendations in Figure \ref{fig:plot_bar_reliance_outcomes}. The results indicate that study participants tend to follow AI recommendations more often when \textit{negative} predicted outcomes are displayed compared to the conditions where no predicted outcomes are displayed (\textit{negative} predicted outcomes: $mean= 80.6\%,\ std = 21.3 \%$; no explanation: $mean= 70.7\%,\ std = 28.4 \%$; feature-based explanation: $mean= 68.4\%,\ std = 24.8 \%$).
This behavior is not observed when predicted outcomes are positive.
Here, reliance is relatively similar across conditions.
Thus, the observed over-reliance for predicted outcomes in general can be largely attributed to an increasing reliance on recommendations to not lend money due to a \textit{negative} predicted outcome. 
This is in line with our hypothesis \textbf{H3}.
In our pilot study, we find a p-value of $p = 0.08$ for the observed difference in reliance across the conditions when AI recommendations are supplemented with negative predicted outcomes.

\subsection{Human-AI decision-making accuracy}\label{sec:AccuracyResults}
In addition to the (over-)reliance behavior of study participants, we analyze the effect of each condition on human-AI decision-making accuracy in general.
These results are summarized in Table~\ref{table:accuracy}.
Our preliminary results indicate that accuracy is not affected by an increasing over-reliance based on predicted outcomes, which is in line with our expectation based on the relationship of reliance and human-AI decision-making accuracy (see hypothesis \textbf{H2}). 
Importantly, the observed human-AI decision-making accuracy in each condition (65.9\% / 65.5\% / 66.9\%) closely resembles $\overline{\mathcal{A}(\mathbf{r})}$ = 66.7\%, \ie, the mean value from the interval defined by our theoretical function $\mathcal{A}(\mathbf{r})$. 
In fact, we observe two compensating effects of reliance on human-AI decision-making accuracy: first, study participants seem to override fewer incorrect AI recommendations that are supplemented with predicted outcomes (29.1\% of incorrect AI recommendations). Second, study participants tend to follow correct AI recommendations including predicted outcomes more often (83.8\%). 
The overall human-AI decision-making accuracy over the three conditions is 66.1\% ($std = 11.0\%$), thus surpassing the accuracy of random guessing (50.0\%).
On average over all three conditions, study participants overrode 35.1\% of incorrect AI recommendations, thus recognizing errors of the AI to a certain degree. 
However, study participants did not always adopt correct AI recommendations (81.6\%), which reduces the overall human-AI decision-making accuracy.
Similar to the previous analysis of reliance, we conduct Kruskal-Wallis tests to evaluate differences in the human-AI decison-making accuracy between conditions.
Here, we find no significant difference in accuracy across the conditions ($p = 0.70$).

\begin{table*}[htbp]
    \centering
    \caption{Observed decision-making accuracy (in \%) by condition.}
    \begin{tabular}{l c c c c}
        \toprule
        \bf Condition & & \bf Overall & \bf AI correct & \bf AI incorrect \\
        & & \bf Mean ($\pm$Std) & \bf Mean ($\pm$Std) & \bf Mean ($\pm$Std) \\
        \midrule
        \bf AI without explanation & & 65.94 ($\pm$9.91) & 79.89 ($\pm$17.97) & 38.04 ($\pm$26.21) \\
        \bf AI with pred. outcomes & & 65.54 ($\pm$9.65) & 83.78 ($\pm$14.98) & 29.05 ($\pm$21.66) \\
        \bf AI with feat.-based expl. & & 66.89 ($\pm$13.49) & 81.58 ($\pm$17.85) & 37.50 ($\pm$23.79) \\
        \midrule
        \bf Average & & 66.11 ($\pm$11.01) & 81.61 ($\pm$17.00) & 35.12 ($\pm$24.28) \\
        \bottomrule
    \end{tabular}
    \label{table:accuracy}
\end{table*}

\section{Discussion and outlook }\label{sec:Discussion}

In our pilot study in the context of peer-to-peer lending, study participants followed correct AI recommendations significantly more often than incorrect ones, regardless of the condition they were assigned to.
Our results thus suggest that study participants were able to recognize when the AI recommendations were incorrect---even when provided with no additional explanation.

\begin{center}
\fbox{
\begin{minipage}{0.8\columnwidth}
    \textbf{Preliminary finding 1:} Across all conditions, study participants were able to distinguish correct from incorrect AI recommendations.
\end{minipage}}
\end{center}

Our results further indicate that study participants tend to be \emph{less} able to distinguish correct from incorrect AI recommendations when AI recommendations are supplemented with predicted outcomes.
This implies that providing predicted outcomes can be detrimental to human-AI decision-making.

\begin{center}
\fbox{
\begin{minipage}{0.8\columnwidth}
    \textbf{Preliminary finding 2:} In contrast to other explanations, predicted outcomes may lead to over-reliance on AI recommendations.
\end{minipage}}
\end{center}

However, we find that over-reliance does not necessarily translate to worse human-AI decision-making performance. In fact, our empirical results indicate that the human-AI decision-making is similar across conditions while reliance levels differ.

\begin{center}
\fbox{
\begin{minipage}{0.8\columnwidth}
    \textbf{Preliminary finding 3:} The empirical human-AI decision-making performance closely resembles the mean of the interval $\overline{\mathcal{A}(\mathbf{r})}$ of the theoretical function $\mathcal{A}(\mathbf{r})$.
\end{minipage}}
\end{center}

We further aim at better understanding potential causes of the observed over-reliance when AI recommendations are supplemented by predicted outcomes. Following prospect theory, we hypothesized that over-reliance is particularly pronounced when predicted outcomes are negative. 

\begin{center}
\fbox{
\begin{minipage}{0.8\columnwidth}
    \textbf{Preliminary finding 4:} The empirical over-reliance observed for predicted outcomes can be largely attributed to a higher reliance on recommendations to not lend money given a negative predicted outcome.  
\end{minipage}}
\end{center}

All our preliminary findings will have to be tested more thoroughly in follow-up work.
As we conducted a pilot study with relatively few study participants, most observed effects are not statistically significant. 
However, we observe several interesting patterns in our results regarding the effects of predicted outcomes on human-AI decision-making that we will investigate in more depth in our main study.
Additionally, we will examine potential reasons for the increase in over-reliance when AI recommendations are supplemented with predicted outcomes.

%
%
%
\bibliographystyle{splncs04}
\bibliography{ecml.bib}

\begin{thebibliography}{10}
\providecommand{\url}[1]{\texttt{#1}}
\providecommand{\urlprefix}{URL }
\providecommand{\doi}[1]{https://doi.org/#1}

\bibitem{Adadi2018}
Adadi, A., Berrada, M.: {Peeking inside the black-box: A survey on explainable
  artificial intelligence (XAI)}. IEEE Access  \textbf{6},  52138--52160 (2018)

\bibitem{Alufaisan2020DoesDecision-Making}
Alufaisan, Y., Marusich, L.R., Bakdash, J.Z., Zhou, Y., Kantarcioglu, M.: Does
  explainable artificial intelligence improve human decision-making? In:
  Proceedings of the AAAI Conference on Artificial Intelligence. vol.~35, pp.
  6618--6626 (2021)

\bibitem{ansari2019prima}
Ansari, F., Glawar, R., Nemeth, T.: {PriMa: A prescriptive maintenance model
  for cyber-physical production systems}. International Journal of Computer
  Integrated Manufacturing  \textbf{32}(4-5),  482--503 (2019)

\bibitem{antoniadi2021current}
Antoniadi, A.M., Du, Y., Guendouz, Y., Wei, L., Mazo, C., Becker, B.A., Mooney,
  C.: Current challenges and future opportunities for {XAI} in machine
  learning-based clinical decision support systems: A systematic review.
  Applied Sciences  \textbf{11}(11), ~5088 (2021)

\bibitem{bastani2020online}
Bastani, H., Bayati, M.: Online decision making with high-dimensional
  covariates. Operations Research  \textbf{68}(1),  276--294 (2020)

\bibitem{Bertsimas2019OptimalTrees}
Bertsimas, D., Dunn, J., Mundru, N.: Optimal prescriptive trees. Journal on
  Optimization  \textbf{1}(2),  164--183 (4 2019)

\bibitem{bertsimas2020prescriptive}
Bertsimas, D., Li, M.L., Paschalidis, I.C., Wang, T.: Prescriptive analytics
  for reducing 30-day hospital readmissions after general surgery. PLOS ONE
  \textbf{15}(9),  e0238118 (2020)

\bibitem{Binns2018ItsDecisions}
Binns, R., Van~Kleek, M., Veale, M., Lyngs, U., Zhao, J., Shadbolt, N.: `{I}t's
  reducing a human being to a percentage': Perceptions of justice in
  algorithmic decisions. In: Proceedings of the 2018 CHI Conference on Human
  Factors in Computing Systems. pp. 1--14 (2018)

\bibitem{Bucinca2020ProxySystems}
Bu{\c{c}}inca, Z., Lin, P., Gajos, K.Z., Glassman, E.L.: {Proxy tasks and
  subjective measures can be misleading in evaluating explainable AI systems}.
  In: Proceedings of the 25th International Conference on Intelligent User
  Interfaces. pp. 454--464 (2020)

\bibitem{Bucinca2021}
Buçinca, Z., Malaya, M.B., Gajos, K.Z.: {To trust or to think: Cognitive
  forcing functions can reduce overreliance on AI in AI-assisted
  decision-making}. Proceedings of the ACM on Human-Computer Interaction
  \textbf{5},  1--21 (2021)

\bibitem{chen2022statistical}
Chen, X., Owen, Z., Pixton, C., Simchi-Levi, D.: A statistical learning
  approach to personalization in revenue management. Management Science
  \textbf{68}(3),  1923--1937 (2022)

\bibitem{Confalonieri2021}
Confalonieri, R., Weyde, T., Besold, T.R., del Prado~Martín, F.M.: {Using
  ontologies to enhance human understandability of global post-hoc explanations
  of black-box models}. Artificial Intelligence  \textbf{296},  103471 (2021)

\bibitem{das2020opportunities}
Das, A., Rad, P.: {Opportunities and challenges in explainable artificial
  intelligence (XAI): A survey}. arXiv preprint arXiv:2006.11371  (2020)

\bibitem{Dodge2019ExplainingJudgment}
Dodge, J., Liao, Q.V., Zhang, Y., Bellamy, R.K.E., Dugan, C.: Explaining
  models: An empirical study of how explanations impact fairness judgment. In:
  Proceedings of the 24th International Conference on Intelligent User
  Interfaces. pp. 275--285 (2019)

\bibitem{Green2019TheMaking}
Green, B., Chen, Y.: {The principles and limits of algorithm-in-the-loop
  decision making}. Proceedings of the ACM on Human-Computer Interaction
  \textbf{3}(CSCW),  1--24 (2019)

\bibitem{guidotti2018survey}
Guidotti, R., Monreale, A., Ruggieri, S., Turini, F., Giannotti, F., Pedreschi,
  D.: {A survey of methods for explaining black box models}. ACM Computing
  Surveys (CSUR)  \textbf{51}(5),  1--42 (2018)

\bibitem{kahneman1979prospect}
Kahneman, D., Tversky, A.: {Prospect theory: An analysis of decision under
  risk}. Econometrica  \textbf{47}(2),  263--292 (1979)

\bibitem{keane2021if}
Keane, M.T., Kenny, E.M., Delaney, E., Smyth, B.: If only we had better
  counterfactual explanations: Five key deficits to rectify in the evaluation
  of counterfactual {XAI} techniques. IJCAI  (2021)

\bibitem{khatri2019analytics}
Khatri, V., Samuel, B.M.: Analytics for managerial work. Communications of the
  ACM  \textbf{62}(4),  100--100 (2019)

\bibitem{kruskal1952use}
Kruskal, W.H., Wallis, W.A.: Use of ranks in one-criterion variance analysis.
  Journal of the American Statistical Association  \textbf{47}(260),  583--621
  (1952)

\bibitem{kuncel2014hiring}
Kuncel, N.R., Klieger, D.M., Ones, D.S.: {In hiring, algorithms beat instinct}.
  Harvard Business Review  (2014)

\bibitem{Lai2019}
Lai, V., Tan, C.: {On human predictions with explanations and predictions of
  machine learning models: A case study on deception detection}. In:
  Proceedings of the Conference on Fairness, Accountability, and Transparency.
  pp. 29--38 (2019)

\bibitem{lakkaraju2017selective}
Lakkaraju, H., Kleinberg, J., Leskovec, J., Ludwig, J., Mullainathan, S.: The
  selective labels problem: Evaluating algorithmic predictions in the presence
  of unobservables. In: Proceedings of the 23rd ACM SIGKDD International
  Conference on Knowledge Discovery and Data Mining. pp. 275--284 (2017)

\bibitem{lee2004trust}
Lee, J.D., See, K.A.: Trust in automation: Designing for appropriate reliance.
  Human Factors  \textbf{46}(1),  50--80 (2004)

\bibitem{lim2019these}
Lim, B.Y., Yang, Q., Abdul, A.M., Wang, D.: {Why these explanations? Selecting
  intelligibility types for explanation goals}. In: IUI Workshops (2019)

\bibitem{mann1947test}
Mann, H.B., Whitney, D.R.: On a test of whether one of two random variables is
  stochastically larger than the other. The Annals of Mathematical Statistics
  pp. 50--60 (1947)

\bibitem{matyas2017procedural}
Matyas, K., Nemeth, T., Kovacs, K., Glawar, R.: A procedural approach for
  realizing prescriptive maintenance planning in manufacturing industries. CIRP
  Annals  \textbf{66}(1),  461--464 (2017)

\bibitem{Miller2019}
Miller, T.: {Explanation in artificial intelligence: Insights from the social
  sciences}. Artificial Intelligence  \textbf{267},  1--38 (2019)

\bibitem{mueller2019explanation}
Mueller, S.T., Hoffman, R.R., Clancey, W., Emrey, A., Klein, G.: Explanation in
  human-{AI} systems: A literature meta-review, synopsis of key ideas and
  publications, and bibliography for explainable {AI}. arXiv preprint
  arXiv:1902.01876  (2019)

\bibitem{naiseh4098528different}
Naiseh, M., Al-Thani, D., Jiang, N., Ali, R.: How different explanations impact
  trust calibration: The case of clinical decision support systems. Available
  at SSRN 4098528  (2022)

\bibitem{nourani2021anchoring}
Nourani, M., Roy, C., Block, J.E., Honeycutt, D.R., Rahman, T., Ragan, E.,
  Gogate, V.: Anchoring bias affects mental model formation and user reliance
  in explainable {AI} systems. In: 26th International Conference on Intelligent
  User Interfaces. pp. 340--350 (2021)

\bibitem{palan2018prolific}
Palan, S., Schitter, C.: {Prolific.ac -- {A} subject pool for online
  experiments}. Journal of Behavioral and Experimental Finance  \textbf{17},
  22--27 (2018)

\bibitem{postma2005improve}
Postma, T.J., Liebl, F.: {How to improve scenario analysis as a strategic
  management tool?} Technological Forecasting and Social Change
  \textbf{72}(2),  161--173 (2005)

\bibitem{Poursabzi-Sangdeh2021ManipulatingInterpretability}
Poursabzi-Sangdeh, F., Goldstein, D.G., Hofman, J.M., Wortman~Vaughan, J.W.,
  Wallach, H.: {Manipulating and measuring model interpretability}. In:
  Proceedings of the 2021 CHI Conference on Human Factors in Computing Systems.
  pp. 1--52 (2021)

\bibitem{rudin2019stop}
Rudin, C.: {Stop explaining black box machine learning models for high stakes
  decisions and use interpretable models instead}. Nature Machine Intelligence
  \textbf{1}(5),  206--215 (2019)

\bibitem{schemmer2022should}
Schemmer, M., Hemmer, P., K{\"u}hl, N., Benz, C., Satzger, G.: {Should I follow
  {AI}-based advice? Measuring appropriate reliance in human-{AI}
  decision-making}. In: ACM CHI '22 Workshop on Trust and Reliance in AI-Human
  Teams (trAIt) (2022)

\bibitem{metaanalysis}
Schemmer, M., Hemmer, P., Nitsche, M., K{\"u}hl, N., V{\"o}ssing, M.: A
  meta-analysis on the utility of explainable artificial intelligence in
  human-{AI} decision-making. arXiv preprint arXiv:2205.05126  (2022)

\bibitem{schemmer2022influence}
Schemmer, M., K{\"u}hl, N., Benz, C., Satzger, G.: On the influence of
  explainable {AI} on automation bias. European Conference on Information
  Systems  (2022)

\bibitem{schoeffer2022relationship}
Schoeffer, J., De-Arteaga, M., Kuehl, N.: On the relationship between
  explanations, fairness perceptions, and decisions. ACM CHI '22 Workshop on
  Human-Centered Explainable AI (HCXAI)  (2022)

\bibitem{townson2020ai}
Townson, S.: {AI can make bank loans more fair}. Harvard Business Review
  (2020)

\bibitem{vereschak2021evaluate}
Vereschak, O., Bailly, G., Caramiaux, B.: How to evaluate trust in
  {AI}-assisted decision making? {A} survey of empirical methodologies.
  Proceedings of the ACM on Human-Computer Interaction  \textbf{5}(CSCW2),
  1--39 (2021)

\bibitem{vossingDesigningTransparencyEffective2022}
V{\"o}ssing, M., K{\"u}hl, N., Lind, M., Satzger, G.: Designing transparency
  for effective human-ai collaboration. Information Systems Frontiers  (May
  2022). \doi{10.1007/s10796-022-10284-3}

\bibitem{vanderWaa2021}
van~der Waa, J., Nieuwburg, E., Cremers, A., Neerincx, M.: {Evaluating XAI: A
  comparison of rule-based and example-based explanations}. Artificial
  Intelligence  \textbf{291},  103404 (2021)

\bibitem{wang2019prescriptive}
Wang, T., Paschalidis, I.C.: Prescriptive cluster-dependent support vector
  machines with an application to reducing hospital readmissions. In: 2019 18th
  European Control Conference (ECC). pp. 1182--1187. IEEE (2019)

\end{thebibliography}

\end{document}